\documentclass{IEEEtran}
\usepackage{cite}
\usepackage{amsmath,amssymb,amsfonts}
\usepackage{textcomp,nicefrac}
\usepackage{xcolor}
\usepackage{graphicx}
\usepackage{subcaption}
\usepackage{lipsum}

\begin{document}

\title{Total Skin Electron Therapy Stanford technique evolution with Monte Carlo
simulation toward personalized treatments for cutaneous lymphoma}

\author{Tullio Basaglia, Patrizia Boccacci, St\'ephane Chauvie, Manuela Chessa,  Daniele D’Agostino, 
Monica Gambaro, Filippo  Grillo Ruggieri, Gabriela Hoff, Maria Grazia Pia, Paolo Saracco, 
Piero Schiapparelli, Giuseppe Scielzo,  Evgueni Tcherniaev and Daniele Zefiro
\thanks{This work was supported in part by Compagnia di San Paolo, Italy.}
\thanks{T. Basaglia is with CERN, Geneva, Switzerland (e-mail: tullio.basaglia@cern.ch)}
\thanks{P. Boccacci, M. Chessa and D. D'Agostino are with DIBRIS, University of Genova, 
and INFN Sezione di Genova  Genova, Italy (e-mail: patrizia.boccacci@unige.it,  daniele.dagostino@unige.it, )}
\thanks{M. Chessa is with DIBRIS, University of Genova, Genova, Italy (e-mail: Manuela.chessa@unige.it)}
\thanks{S. Chauvie is with Santa Croce e Carle Hospital, Cuneo, Italy (e-mail: chauvie.s@ospedale.cuneo.it)}
\thanks{M. Gambaro and P. Schiapparelli are with Galliera Hospital, Genova, Italy (e-mail: monica.gambaro@galliera.it and piero.schiapparelli@galliera.it)}
\thanks{F. Grillo Ruggieri is with the Department of Radiation Oncology, Proton Therapy Center at Paul Scherrer Institute, Villigen, Switzerland; he was with the Department of Radiation Oncology, Galliera Hospital, Genoav, Italy  (e-mail: f.grilloruggieri@gmail.com)}
\thanks{G. Hoff, M. Grazia Pia and P. Saracco are with INFN Sezione di Genova, Genova, Italy (e-mail: gabriela.hoff@ge.infn.it, mariagrazia.Pia@ge.infn.it and paolo.saracco@ge.infn.it)}
\thanks{G. Scielzo is with Policlinico di Monza, Medical Physics Consulant, Monza, Italy (e-mail: giuseppescielzo@gmail.com)}
\thanks{E. Tcherniaev is with Tomsk State University, Tomsk, Russia and CERN, Geneva, Switzerland (e-mail: Evgueni.Tcherniaev@cern.ch)}
\thanks{D. Zefiro is with the IRCCS AOU Policlinico San Martino (e-mail: daniele.zefiro@hsanmartino.it)}}
\maketitle

\begin{abstract}
Current Total Skin Electron Therapy (TSET) Stanford technique for cutaneous
lymphoma, established in the 70's, involves a unique irradiation setup, i.e.
patient's position and beam arrangement, for all patients with ensuing great
variability in dose distribution and difficult dose optimization.
A Geant4-based simulation has been developed to explore the possibility of
personalizing the dose to each patient's anatomy.
To achieve this optimization of the treatment method, this project enrolls
different aspects of the clinical and computational techniques: starting with
the knowledge of the experimental parameters involving TSET practice, passing
through an innovative approach to model the patient's anatomy, a precise
description of the electron beam and a validated configuration of the physics
models handling the interactions of the electrons and of secondary particles.
The Geant4-based simulation models the patient as a tessellated solid derived
from the optical scan of her/his body, realistically reproduces the irradiation
environment in detail and calculates the energy deposition corresponding to each
facet of the patient's scanned surface.
The resulting three-dimensional dose distribution constitutes the basis for the
personalization of the medical treatement as appropriate to each patient's
specific characteristics.
\end{abstract}

\begin{IEEEkeywords}
Geant4, simulation, Total Skin Electron Therapy, computational dosimetry
\end{IEEEkeywords}

\section{Introduction}

Total Skin Electron Therapy (TSET) is a special radiotherapy technique to treat
cutaneous lymphoma.
It is commonly based on a method originally developed at
Stanford in the eighties of the past century~\cite{b1, b2, b3, b4, b5}:
the standing patient is exposed at several angled orientations to large
electrons fields obtained using a 6~MeV linear accelerator.
Under this conditions it is assumed that the delivered dose distribution of a
few millimeters in depth is sufficiently homogeneous to cover the tumor
infiltration of cutaneous lymphoma.

The treatment consists of a single standard technique for all patients: in the
application of this technique, geometrical fields setup, dose contribution, and
patient position are fixed, resulting in standard non-modifiable dose
distributions.
Different degrees of tumor infiltration and extension, patient size, and shape
variability among different patients are not taken into account in current
practice.
This is the opposite of what happens with other radiotherapy techniques, where
the successful trend is the adaptation of every single treatment to
patient-specific normal and disease anatomy.

The possibility to adapt and personalize the treatment for TSET is related to
the development of fast and reliable calculation procedures of the skin dose
distribution made on patient-specific surface anatomy.
In this way, it would be possible to transform this technique from a ``one size
fit all'' therapy to a finely personalized treatment, adapted to subtle
differences in the disease involvement - e.g., in the trunk rather than in the
arms or head and other combinations,  also  planning different dose contributions
to the upper or lower body part and/or surface.

To address this challenging dose calculation requirements, a Monte Carlo 
simulation approach has been devised in the context of the SkinScan project.
The peculiarities of this radiotherapy technique required innovative modeling
methods of the patient, which, to the best of our knowledge, are applied in a
radiotherapy simulation for the first time.
They also required rigorous validation of the physics modeling involved in the
simulation, to ensure that it would produce reliable dose distributions in thin
layers corresponding to the patient's skin.

This conference paper highlights the main features of the R\&D project and reports 
the first results.

\section{Simulation}

The Monte Carlo simulation of TSET was developed using the Geant4~\cite{g4nim,
g4tns, g4nim2} toolkit, version 11.0p01.
It faithfully reproduces a real-life irradiation environment at the Galliera
Hospital in Genova, Italy and adopts an innovative approach to realistically
represent the patient.

\subsection{Model of the TSET setup} 

\begin{figure}[htbp]
\centerline{\includegraphics[width=8.5cm]{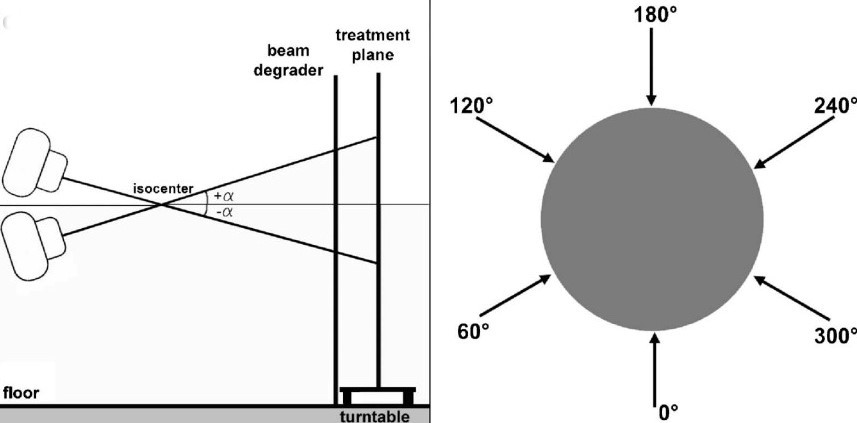}}
\caption{Scheme of Total Skin Electron Therapy, as currently applied according
to the guidelines of the Stanford technique. Lateral view (left) illustrating the
Source to detector Surface Distance (SSD) of 353~cm to the treatment plane, a
beam degrader placed at 320~cm from the source and the two beams, with polar
angle $\alpha$ = 19${^\circ}$. Top view (right) showing the six beam angles
around the patient to perform the complete treatment. Figures 1(c) and 1(d)
from~\cite{b5}.}
\label{figScheme}
\end{figure}

The simulation models the real-life setup described in ~\cite{b5};
a schematic view is illustrated in Fig. \ref{figScheme}.

The TSET irradiation configuration involves six dual beams of electrons produced
by a 6~MeV clinical linear accelerator, which encompass the entire body height
of a standardized patient in vertical position.

The simulation initially implemented a simplified model of the electron beam
delivered by a Varian 2100 C$/$D linear accelerator, equipped with the special
procedure for HDTSe-high dose rate total skin electron mode (6~MeV).
An investigation to model primary electrons using the IAEA database for external
radiotherapy \cite{iaeabeams} is in progress.
The primary electrons are degraded through the passage in air and by a plastic screen
from the initial energy of 6 MeV down to about 2 MeV at the surface of the skin.

\subsection{Model of the patient anatomy}

The project developed a novel computational approach to the patient's geometry:
it creates a 3-dimensional model of the body surface from an optical scan of the
patient.
The coordinates of the scanned data points on the patient's surface define a
numerical lattice, which is the basis to build the geometry of a tessellated solid.

This innovative computational solution overcomes the drawbacks of conventional
anthropomorphic phantom models derived from CT (computed tomography) scans: it
efficiently satisfies the peculiar TSET requirements of precisely modeling the
patient's surface, while avoiding the unnecessary computational burden of
internal anatomy details.

The requirements of this novel approach motivated improvements to the
functionality and the computational performance of Geant4 tessellated solids.
A major requirement concerned the ability to score observables of dosimetric
interest in identifiable components of tessellated geometries; it was satisfied
by the development of additional capabilities in Geant4 tessellated solid.

\subsection{Physics modeling and validation}

The simulation requires rigorous experimental validation to assess its
reliability in a sensitive application context such as TSET.
For this purpose, comparisons have been made with experimental data, 
both at the level of Geant4 physics components
and at the macroscopic level with dosimetric measurements.

A thorough validation test was performed to identify the state of the art among
all the Geant4 physics modeling options relevant to the simulation of the energy
deposited in thin layers of materials  by electrons in the energy range relevant to 
TSET~\cite{sandia_nss2022}.

The complete simulation application was validated using regular quality control
dosimetric measurements and experimental data specific to the characterization
of the TSET technique reported in ~\cite{b5}.

The validation process exploits rigorous methods of statistical inference.

\subsection{Dosimetry}

Geant4 provides a variety of means to evaluate the energy deposited by primary
and secondary particles in sensitive volumes, i.e., in the patient in the TSET
scenario.
In a preliminary approximation, electrons are assumed to deposit all their energy
in correspondence to the facet of the patient's tessellated solid where they enter.
Further evaluations are in progress to optimize the estimate of the energy deposition 
in the patient, considering both the precision and the computational performance of the simulation.

\section{Preliminary Results}

Some preliminary results of the simulation are illustrated here.

\begin{figure}
\begin{subfigure}[b]{0.5\columnwidth}
        \centering
        \includegraphics[width=4.1cm]{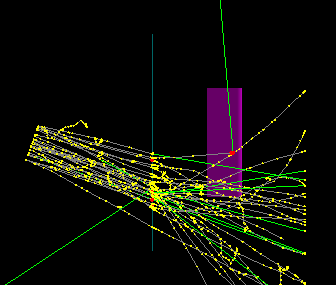}
        \caption{Lateral view bottom irradiation}
    \end{subfigure}%
\begin{subfigure}[b]{.5\columnwidth}
        \centering
        \includegraphics[width=4.1cm]{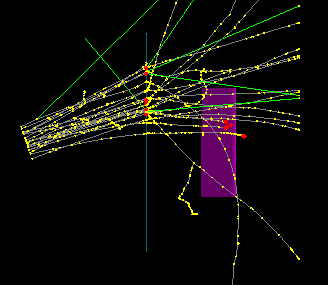}
        \caption{Lateral view top irradiation}
    \end{subfigure} \\%
\begin{subfigure}[b]{.99\columnwidth}
        \centering
        \includegraphics[width=8.5cm]{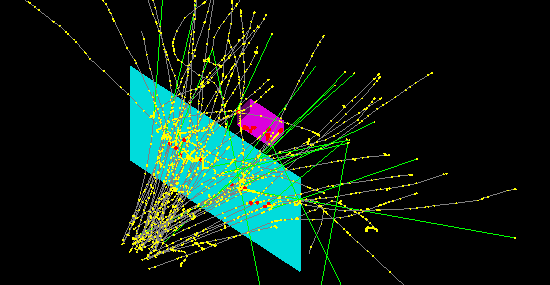}
        \caption{Angled view from dual beam irradiation}
    \end{subfigure}%
\caption{Illustration of the dual beam irradiation configuration showing  (a)  bottom surface irradiation, (b)  top surface irradiation and (c) an angled view of the dual beam irradiation configuration. This is a simplified geometry: the beam degrader is represented as a cyan thin parallelepiped volume and the phantom
as a cylindrical volume in pink. The electron tracks are shown in yellow and the photon tracks in green.}
\label{figIrradSimul}
\end{figure}

The  dual beam irradiation configuration modeled in the simulation 
is shown in Fig. \ref{figIrradSimul}; in this picture, the target is a simple
cylinder of water.

\begin{figure}
\begin{subfigure}[b]{0.5\columnwidth}
        \centering
        \includegraphics[width=4.3cm]{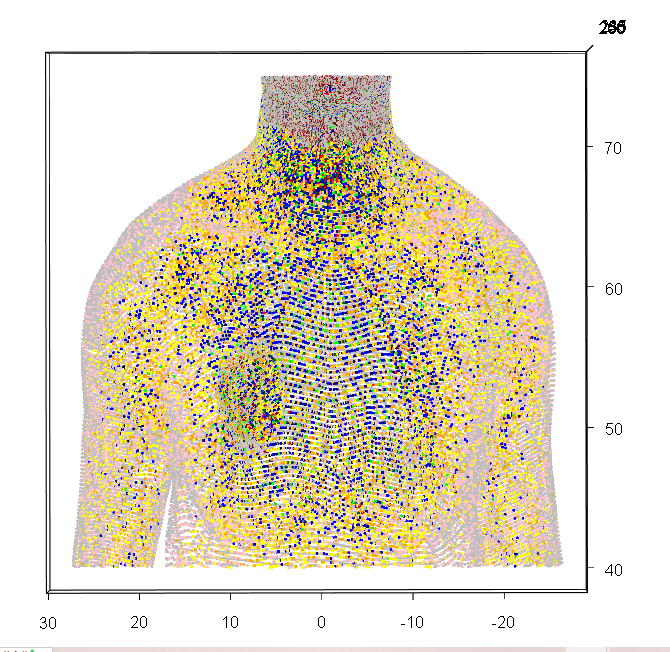}
        \caption{Frontal view}
    \end{subfigure}%
\begin{subfigure}[b]{.5\columnwidth}
        \centering
        \includegraphics[width=4.3cm]{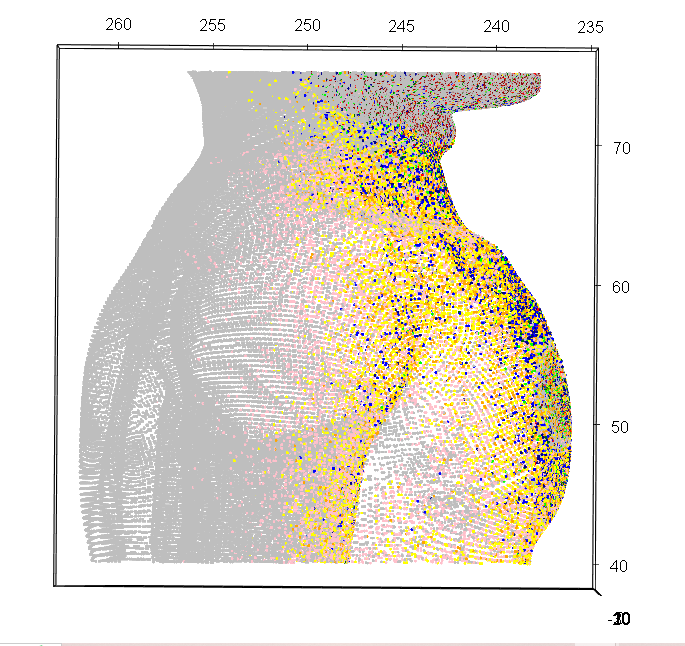}
        \caption{Lateral view}
    \end{subfigure} \\%
\begin{subfigure}[b]{.5\columnwidth}
        \centering
        \includegraphics[width=4.5cm]{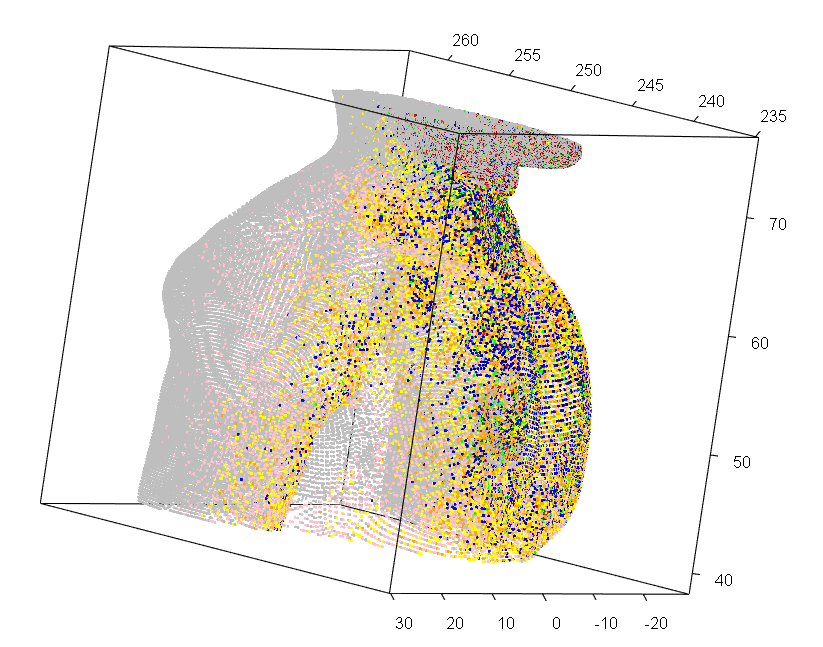}
        \caption{Angled view}
    \end{subfigure}%
\begin{subfigure}[b]{.5\columnwidth}
        \centering
        \includegraphics[width=4.1cm]{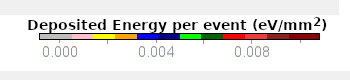}
        \caption{Color scale the normalized deposited energy}
    \end{subfigure}%
\caption{Three-dimensional map of the normalized deposited energy for the
tessellated anthropomorphic phantom for frontal dual beam irradiation of the patient 
in the TSET Stanford technique:
(a) frontal view, (b) lateral view (c) angled view and (d) the color scale of the three-dimensional maps.}
\label{fig3Dmap}
\end{figure}

Fig. \ref{fig3Dmap} illustrates the simulated three-dimensional deposited energy in a
frontally irradiated phantom, represented by a tessellated solid derived from
the optical scan of a man's surface.
It focuses on an anatomical part where the achievement of unifom dose distribution
is especially difficult due to the complexity of irradiating areas that involve
maximal changes of the body shape and skin surface orientation to the electron
beam.
The map represents the distribution of the energy deposition associated with the
facets of the tessellated solid, normalized to the surface of each facet of the
solid, consistent with a defined the colour scale.

The ability of the simulation to quantify and localize the energy deposition,
and consequently the dose distribution, in a realistic geometrical model of each
patient allows the optimization of the irradiation according to the specific
characteristics of each patient.

\section{Conclusions}

The peculiarities of the TSET radiotherapy technique involved the development of
innovative computational solutions in the context of the SkinScan project to
perform a realistic simulation of the irradiation scenario.

The patient modeling method and its computational realization through a tessellated solid 
are applied in a radiotherapy simulation for the first time,  to the best of our knowledge.
The ability to calculate the energy deposited by facet of the phantom required 
the development of additional functionality in Geant4 tessellated solid.

Preliminary results demonstrate the capability to simulate three dimensional
maps of realistic dose distributions in patient-specific geometries, which open
the path towards the personalization and optimization of TSET.

The Geant4-based simulation developed by the SkinScan project represents a
significant contribution to fully personalized treatment in TSET practice.
Detailed description of the SkinScan software system and simulation application results
will be documented in forthcoming journal publications.

\section*{Acknowledgment}

The authors thank the Computing Service and the Administration of the INFN Section of Genova for
valuable support, and the INFN National Computational Services for providing fundamental infrastructure 
for the software development.
The CERN Library has provided substantial assistance and essential reference material for this research.

\end{document}